\documentclass[10pt]{blois07}
\usepackage{graphicx}

\usepackage{cite,./mcite}

\newcommand{\be}{\begin{equation}}
\newcommand{\ee}{\end{equation}}
\newcommand{\beeq}{\begin{eqnarray}}
\newcommand{\eeeq}{\end{eqnarray}}

\def\funp{{I\!\!P}}
\def\xp{x_{{I\!\!P}}}

\def\qbar{\overline{q}}

\def\gev{\mbox{\rm GeV}}

\def\eto{{\rm e}}

\newcommand{\eq}{\!\!\!&=&\!\!\!}

\begin{document}
\title{Diffractive structure function  $F_L^D$ from fits with higher twist}
\author{Krzysztof Golec-Biernat$^{\,1,2}$\protect\footnote{~~talk presented at EDS07}~,
Agnieszka \L{}uszczak$^{\,1}$}
\institute{$^1$Institute of Nuclear Physics Polish Academy of 
Sciences, Cracow, Poland\\ 
$^2$Institute of Physics, University of  Rzesz\'ow, Rzesz\'ow, Poland}
\maketitle
\begin{abstract}
We make predictions for the diffractive longitudinal structure function $F_L^D$
to be measured at HERA, based on DGLAP based fits of diffractive parton 
distributions with  twist--4 contribution. This contribution describes
diffractive $q\bar{q}$ production from longitudinal photons and  significantly changes
predictions for $F_L^D$ obtained in pure DGLAP analyses.
\end{abstract}

\section{Introduction}
\label{sec:1}

We are interested in diffractive deep inelastic scattering (DDIS) at HERA which  provide a very interesting example of processes with a clear interplay
between hard and soft aspects of QCD interactions. In these processes,
a diffractive system is formed which is separated in rapidity from the scattered proton.
The most important observation made at HERA is that diffractive processes in DIS are not rare
and constitute up to $15\%$ of  all deep inelastic scattering events
\cite{Chekanov:2004hy, Chekanov:2005vv, Aktas:2006hx, Aktas:2006hy}.

After integration over the proton azimuthal angle, the diffractive cross section  is given in terms two structure functions, $F_2^D$ and $F_L^D$, which depend on four variables:
Bjorken-$x$, photon virtuality $Q^2$ and two additional variables
\be
\xp=\frac{Q^2+M^2-t}{Q^2+W^2}\,,~~~~~~~~~~~~~~t=(p-p^\prime)^2\,.
\ee 
Here $M$ is  mass of the diffractive system, $W$ is invariant energy of the gamma-proton system and $p,\,p^\prime$ are incident and scattered proton momenta.
In our analysis, the diffractive structure functions are  decomposed into the leading
twist--2  and higher twist contributions
\be\label{eq:5}
F_{2,L}^D\,=\,F_{2,L}^{D({tw}2)}
\,+\,{F_{L}^{D({tw}4)}}\,+\,\ldots\,.
\ee
In the Bjorken limit, the leading  part depends logarithmically on 
$Q^2$  while the twist--4 part is suppressed by additional power of $1/Q^2$. 
However, this contribution dominates over the twist--2 one 
for small diffractive masses, $M^2\ll Q^2$, playing especially important role in DDIS.  
Physically, the twist--4 contribution is given by diffractively produced $q\bar{q}$ pairs from  {\it longitudinally} polarized virtual photons.  The effect of this contribution
is particularly important for the diffractive longitudinal structure function $F_L^D$ which is supposed to be determined from the HERA data.

\section{Twist--2 contribution}
\label{sec:2}

The twist--2 contribution is given in terms of diffractive parton distributions (DPD) through standard collinear factorization formulae
\cite{Trentadue:1993ka,Berera:1994xh,Collins:1994zv,Berera:1995fj}. 
In the next-to-leading logarithmic approximation we have
\beeq
\label{eq:6a}
F_{2}^{D({tw}2)}(x,Q^2,\xp,t) \eq 
S_D + \frac{\alpha_s}{2\pi}\left\{C^S_{2}\otimes S_D + C^G_{{2}}\otimes G_D\right\}
\\\nonumber
\\\label{eq:6b}
F_{L}^{D({tw}2)}(x,Q^2,\xp,t) \eq ~~~~~~~~~~
\frac{\alpha_s}{2\pi}\left\{C^S_{L}\otimes S_D + C^G_{{L}}\otimes G_D\right\}
\eeeq
where $\alpha_s$ is the strong coupling constant and $C_{2,L}^{S,G}$ are coefficients functions, known from inclusive DIS \cite{Furmanski:1980cm,Furmanski:1981cw}.   The integral convolution is performed for the  longitudinal momentum fraction and reads
\be
(C\otimes F)(\beta)= \int_\beta^1 dz\,  C\left({\beta}/{z}\right) F(z) \,.
\ee
Notice that in the leading order, when terms proportional to $\alpha_s$ are neglected, $F_L^{D(tw2)}=0$. The functions $S_D$ and $G_D$ are built from diffractive quark $(q_D^f)$
and gluon $(g_D)$ distributions
\be\label{eq:7}
S_D(x,Q^2,\xp,t) \,=\, 
\sum_{f=1}^{N_f} e_f^2\,\beta\left\{q^f_D(\beta,Q^2,\xp,t)+
\overline q^f_D(\beta,Q^2,\xp,t)\right\}
\ee
and 
\be
G_D(x,Q^2,\xp,t)=\beta g_D(\beta,Q^2,\xp,t)\,.
\ee
From these equations we see that  the new variable 
\be
\beta=\frac{x}{\xp}=\frac{Q^2}{Q^2+M^2}
\ee
plays the role of the Bjorken variable in DDIS.

The DPD evolve with $Q^2$ with  the DGLAP evolution equations \cite{Collins:1998rz}
for which $(\xp,t)$ are external parameters. In this analysis we
assume {\it Regge factorization} for these variables:
\beeq
\label{eq:reggefac}
q^f_D(\beta,Q^2,\xp,t)\eq\,f_\funp(\xp,t)\,\,q_\funp^f(\beta,Q^2)
\\
g_D(\beta,Q^2,\xp,t)\eq\,f_\funp(\xp,t)\,\,g_\funp(\beta,Q^2)\,.
\eeeq
The motivation for such a factorization is a model of diffractive interactions with the pomeron exchange
\cite{Ingelman:1984ns}. In this model $f_\funp$ is the pomeron flux
\be\label{eq:pomflux}
f_\funp(\xp,t)\,=\,N\,\frac{F^2_\funp(t)}{8\pi^2}\,{\xp^{1-2\,\alpha_\funp(t)}}\,,
\ee
where $\alpha_\funp(t)=\alpha_{\funp}(0)+\alpha_{\funp}^\prime\/ t$ is the pomeron  Regge trajectory and 
\be\label{eq:formfactor}
F^2_\funp(t)\,=\,F^2_\funp(0)\,\eto^{-B_D |t|}\,,
\ee
is elastic formfactor which  describes the pomeron--proton coupling.
The diffractive slope is taken form HERA data, $B_D=5.5~\gev^{-2}$,
and $F^2_\funp(0)=54.4~\gev^{-2}$ \cite{Collins:1994zv}. For the pomeron trajectory,  $\alpha_{\funp}^\prime=0.25~\gev^{-2}$ but the intercept $\alpha_{\funp}(0)$ is fitted to the analyzed data. The diffractive quark  distributions are flavour independent, thus
\be
q_\funp^f(\beta,Q^2)\,=\,\overline{q}_\funp^f(\beta,Q^2)\,\equiv\,
\frac{1}{2N_f}\,\Sigma_\funp(\beta,Q^2)
\ee
where $N_f$ is a number of active flavours and $\Sigma_\funp$ is singlet distribution.
We fit $\Sigma_\funp(\beta,Q^2)$ and $G_\funp(\beta,Q^2)$ at an
initial scale $Q^2_0=1.5~\gev^2$ to diffractive data from HERA, using
the DGLAP evolution equations in the next-to-leading order approximation.
We also include charm quark contribution into the analysis \cite{Golec-Biernat:2007hx}.
\begin{figure}[t]
\begin{center}
\includegraphics[width=6cm]{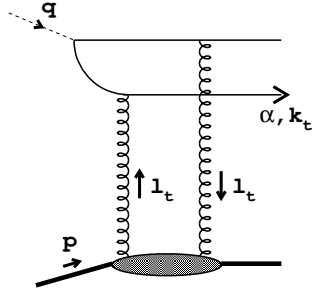}
\caption{Twist-4 contribution from longitudinally polarized photons. Two  gluons
here model the  pomeron exchange which is later unitarized and  effectively described
by the dipole cross section, see the text below.}
\label{fig:0}
\end{center}
\end{figure}
\section{Twist--4 contribution}
\label{sec:3}

The twist--4 contribution describes  diffractive production of the 
$q\overline{q}$ pairs from  longitudinally polarized virtual photons, see Fig.~\ref{fig:0}. Although formally suppressed
by $1/Q^2$, this contribution dominates over the twist--2 contribution for small diffractive masses, $M^2\ll Q^2$ (or $\beta\to 1$) \cite{Wusthoff:1997fz,Bartels:1998ea,Golec-Biernat:1999qd}. This is why it cannot be neglected in the analysis of diffractive DIS data.

We used the following form of the twist--4 contribution which has to be added to the diffractive structure functions $F_2^{D(tw2)}$ and $F_L^{D(tw2)}$ \cite{Golec-Biernat:2001mm}:
\be
\label{eq:flqq}
F_{Lq\bar{q}}^{D(tw4)}=
\frac{3}{16\pi^4\xp}\,\eto^{-B_D|t|}\,\sum_f e_f^2\,
\frac{\beta^3}{(1-\beta)^4}\;
\int\limits_0^{\frac{Q^2(1-\beta)}{4\,\beta}} \!dk^2\
\frac{\displaystyle {k^2}/{Q^2}}
{\displaystyle \sqrt{1-\frac{4\beta}{1-\beta}\frac{k^2}{Q^2}}}\,
\phi_0^2
\ee
with
\be
\label{eq:phi1}
\phi_0
\;=\;k^2
\int\limits_0^\infty dr\, r\, K_{0}\!\left(\sqrt{\frac{\beta}{1-\beta}}kr\right)
J_{0}(kr)\,  \hat{\sigma}(\xp,r)
\ee
where $K_0$ and $J_0$ are Bessel functions.  Strictly speaking, formula
(\ref{eq:flqq})
contains all powers of $1/Q^2$ but the twist--4 part, proportional to $1/Q^2$, dominates.
The function $\hat{\sigma}(\xp,r)$ in eq.~(\ref{eq:phi1})
is the dipole--proton cross section which describes the interaction of a
quark--antiquark dipole of transverse size $r$ with the proton. Following \cite{Golec-Biernat:1998js}, we choose
\be\label{eq:dipcs}
\hat{\sigma}(\xp,r)\,=\,\sigma_0\,\{1-\exp{(-r^2Q_s^2(\xp))}\}
\ee
where $Q_s^2(\xp)=(\xp/x_0)^{-\lambda}~{\rm GeV}^2$ is a saturation scale which provides  energy dependence of the twist--4 contribution. The dipole cross section parameters,
$\sigma_0=29~{\rm mb},~x_0=4\cdot10^{-5}$ and $\lambda=0.28$,
are taken from  \cite{Golec-Biernat:1998js} (Fit 2 with charm).
This form  of the dipole cross section provides successful description of 
the inclusive and  diffractive data from HERA.

In addition to the twist terms, 
we also consider a reggeon contribution, described in detail in 
\cite{Golec-Biernat:2007hx}, which improves  fit quality through better
dependence on $\xp$.

\section{Fit results}
\label{sec:4}

In our analysis, we use diffractive data on $F_2^D$ (or reduced cross section
$\sigma_r^D$) from the H1 \cite{Aktas:2006hx,Aktas:2006hy} and ZEUS \cite{Chekanov:2004hy,Chekanov:2005vv} collaborations.
These data were obtained in different
kinematical regions, using different methods of their analysis, thus
we decided to analyse them separately. For each data set we performed two fits: 
with the twist--4 present and without this term (pure DGLAP fits).
In this way,  we obtained two sets of diffractive
parton distributions which allow us to make predictions for the longitudinal structure function $F_L^D$. A full discussion of fit details is given in \cite{Golec-Biernat:2007hx}.

\begin{figure}[t]
\begin{center}
\includegraphics[width=12cm]{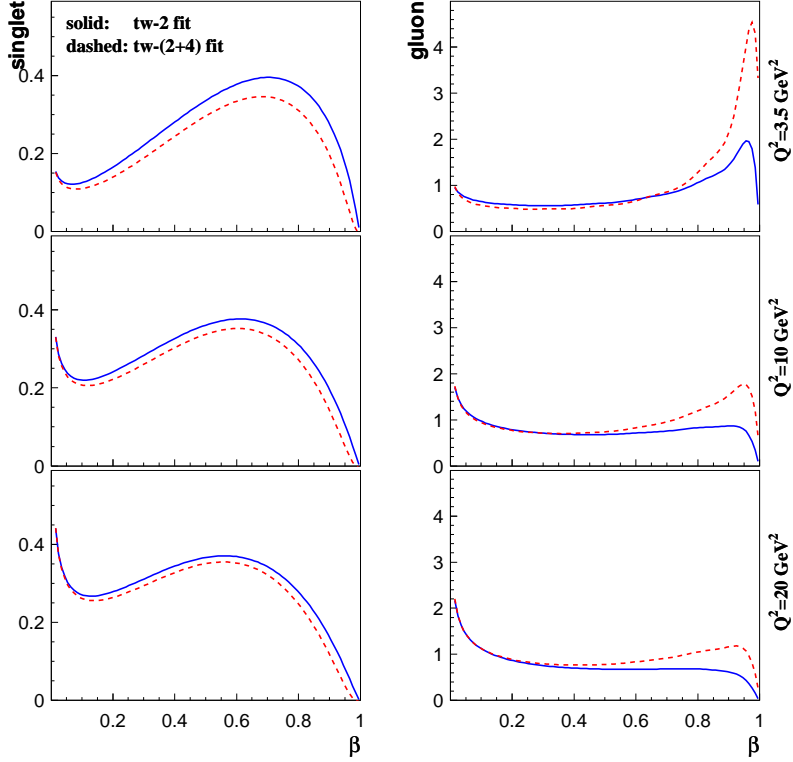}
\caption{Pomeron parton distributions: quark singlet  (left) and gluon   (right) from the H1 data. Solid lines: twist--2 fit; dashed lines: fit with twist--4 term.}
\label{fig:2}
\end{center}
\end{figure}

The diffractive parton distributions  (DPD) from the analysis of H1 data
are shown in  Fig.~\ref{fig:2}. They  are given in terms of the pomeron parton distributions which can be multiplied by the pomeron flux $f(\xp,t)$ to obtain the DPD.
We see   that the singlet quark distributions
from the two discussed fits are practically the same, while the gluon distributions  are different. The gluon from the fit with twist--4 is stronger peaked near $\beta\approx 1$ than in the twist--2 fit.
This somewhat surprising result can be understood by analyzing the logarithmic slope
$\partial F_2^D/\partial \ln Q^2$ for fixed $\beta$. 
In the leading logarithmic approximation, we have from the DGLAP equations
\be
\frac{\partial F_2^D}{\partial \ln Q^2}\sim\frac{\partial\Sigma_\funp}{\partial \ln Q^2}
=P_{qq}\otimes \Sigma_\funp \,+\, P_{qG}\otimes G_\funp\,-\,\Sigma_\funp \int P_{qq}
\ee
where the negative term sums virtual corrections. For large $\beta$, the measured slope 
is negative which means that the negative term must dominate over the positive terms describing real emissions. 
The addition of the  twist--4 contribution to $F_2^D$, proportional to $1/Q^2$,  contributes  negative value to the slope which has to be compensated by a larger gluon distribution near $\beta\approx 1$ in order to describe the same data.

\begin{figure}[t]
\begin{center}
\includegraphics[width=12cm]{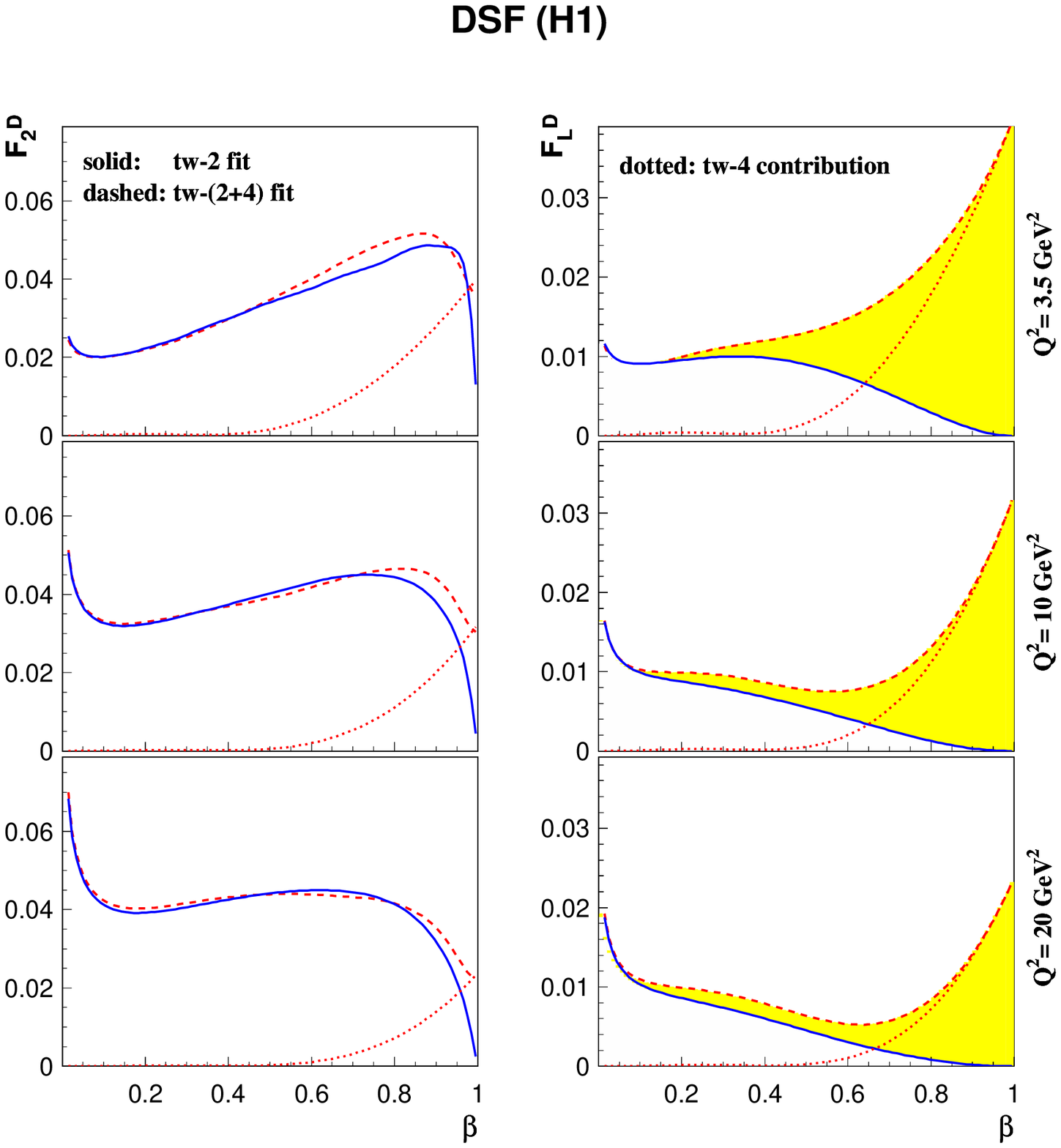}
\caption{Diffractive structure functions $F_2^D$ (left) and $F_L^D$ (right) from fits to the H1 data for $\xp=10^{-3}$. Solid lines: twist--2 fit; dashed lines: twist--(2+4) fit; dotted lines: twist--4 contribution.}
\label{fig:3}
\end{center}
\end{figure}

In Fig.~\ref{fig:3}, we show the diffractive structure functions
resulting from the determined parton distributions. As expected,  $F_2^D$ is practically the same in both fits.
However,  the $F_L^D$ curves are significantly different due to the twist--4 contribution
(shown as the dotted lines). Let us emphasised that both sets of curves were
found in  the fits which well describe the existing data, especially  in the region of large  $\beta$ where twist--4 is important.
Thus, an independent {\it measurement} of $F_L^D$ in this region would be 
very important for confirmation of the QCD approach to diffraction. 

\begin{figure}[t]
\begin{center}
\includegraphics[width=12cm]{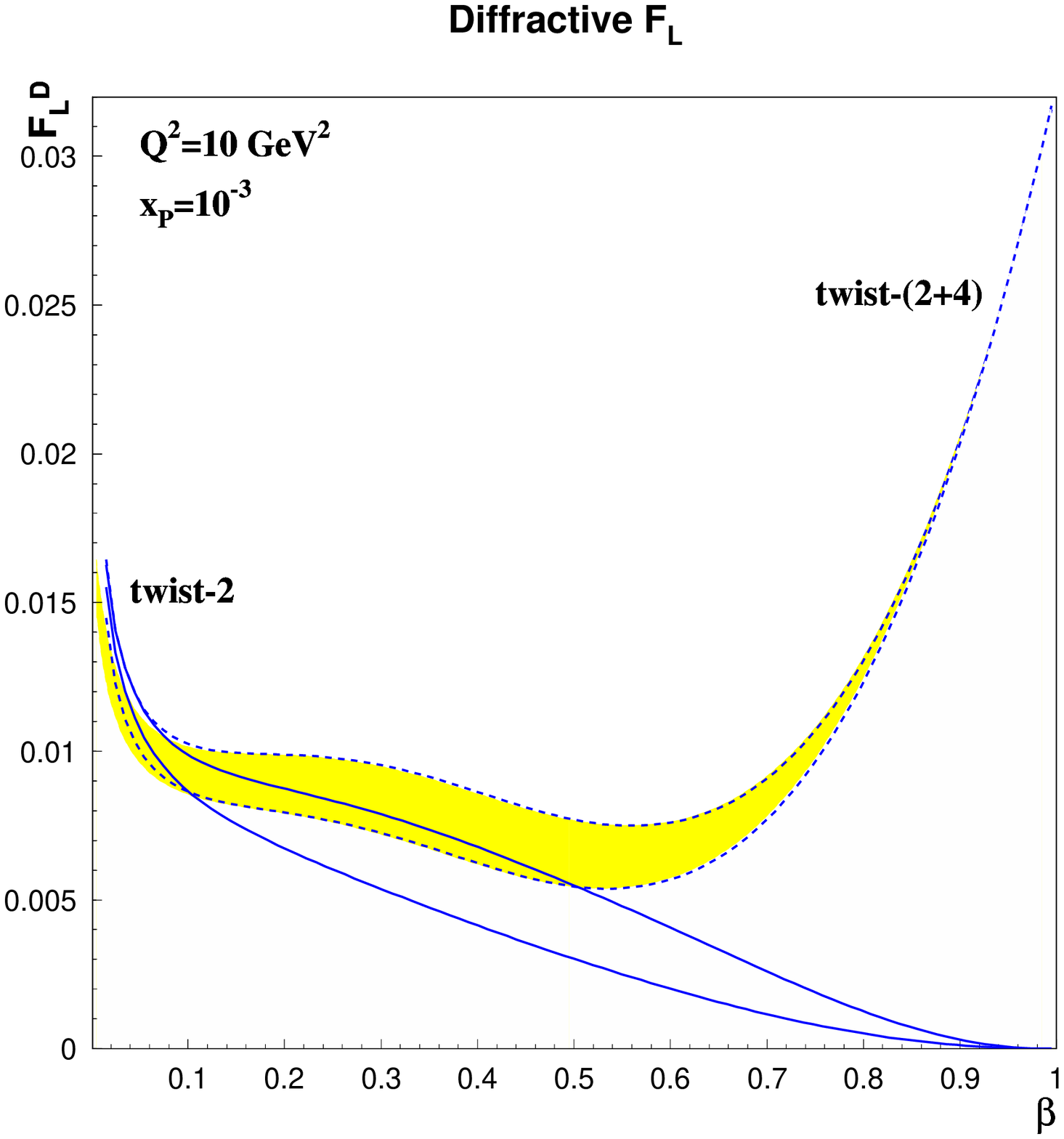}
\caption{Predictions for $F_L^{D(3)}$ for $\xp=10^{-3}$ and $Q^2=10~{\rm GeV^2}$ 
from fits with twist--4 to the H1 (upper dashed line) and ZEUS (lower dashed line) data. The solid lines show predictions from pure twist--2 fits to the same data: from
H1 (upper) and ZEUS (lower).}
\label{fig:4}
\end{center}
\end{figure}

We summarize the effect of the twist--4 contribution in Fig.~\ref{fig:4}, showing
the predictions for the diffractive longitudinal structure function. Ignoring this contribution, we find the two solid curves
coming from the pure twist--2 analyses of the H1 (upper) and ZEUS (lower) data.
With twist--4, the dashed curves are found, the upper curve from the H1 data and the lower one from the ZEUS data analyses. There is a significant difference between these predictions in the region of large
$\beta$. We believe that this effect will be confirmed by the forthcoming  analysis of the HERA
data.


\section{Summary}
\label{sec:5}

We performed the analysis of the diffractive data from HERA, determining diffractive parton distributions. In addition to the standard twist--2 formulae, we also
considered the twist--4 contribution. This contribution comes from the $q\qbar$ diffractive production from  longitudinally polarized virtual photons and dominates for
$M^2\ll Q^2$ (large $\beta$).
The twist--4 contribution leads to the diffractive gluon distribution which is 
stronger peaked at $\beta\approx 1$ than the gluon distribution from the pure twist--2 fits.

The main result of our analysis is a new prediction for 
the longitudinal diffractive structure function $F_L^D$. The twist--4 term 
significantly enhances $F_L^D$ in the region of $\beta> 0.6$.
A measurement of this function at HERA in the region of large $\beta$ should confirm the presented expectations which are based on  perturbative QCD calculations.

\newpage
\centerline{ACKNOWLEDGEMENTS}

We would like to acknowledge  support from  the MEiN research grant~
1~P03B~028~28 (2005-08),  the Research Training Network HEPTools  (MRTN-2006-CT-035505) and from  the Polish-German Joint Project "Hadronic final states and parton distribution functions".

\begin{footnotesize}
\bibliographystyle{blois07} 
{\raggedright
\bibliography{mybib}
}
\end{footnotesize}
\end{document}